\documentclass[prd,aps,twocolumn,preprintnumbers, showpacs, nofootinbib,superscriptaddress,notitlepage]{revtex4-1}
\usepackage{amssymb,amsthm,amsmath}
\usepackage{graphicx}   
\usepackage{color}      
\usepackage{slashed}    
\usepackage{verbatim}
\usepackage[normalem]{ulem}
\usepackage{rotating}   
\usepackage{multirow}   
\begin{document}
\title{ A novel method for searching the $\Xi_c^{0/+}$-$\Xi_c^{\prime 0/+}$ mixing effect in the angular distribution analysis of a four-body $\Xi_c^{0/+}$ decay}

\author{  Zhi-Peng Xing}\email{Email:zpxing@sjtu.edu.cn}
\affiliation{Tsung-Dao Lee Institute, Shanghai Jiao Tong University, Shanghai 200240, China}

  \author{ Yu-Ji Shi} \email{Email:shiyuji92@126.com}
  \affiliation{ Helmholtz-Institut f\"ur Strahlen- und Kernphysik and Bethe Center 
  for Theoretical Physics,\\ Universit\"at Bonn, 53115 Bonn, Germany\\}

\begin{abstract}
In this work, we raised a novel method for searching the $\Xi^{0+}_c$-$\Xi_c^{0+\prime}$ mixing effect in an angular distribution analysis of the $\Xi_c\to\Xi^{(\prime)}(\Lambda \pi)\ell^+\nu$ decay, where the mixing effect can be observed by the appearance of the $\Xi^{\prime}$ resonant.  Armed with this angular distribution,  the decay branching fraction and the forward-backward asymmetry are predicted. We pointed out that the forward-backward asymmetry, as a function of the invariant mass square of $\Xi^{(\prime)}$ and the $\Xi^{0+}_c$-$\Xi_c^{0+\prime}$ mixing angle $\theta_c$,  can be used to distinguish the two resonants $\Xi^{(\prime)}$ and even provide a possibility to  determine the exact mixing angle.
\end{abstract}

\maketitle

\section{Introduction}
Particle physics describes all the fundamental materials and interactions in our universe.  After 2012, all the elementary particles predicted by the Standard Model (SM) had been observed \cite{ATLAS:2012yve,CMS:2012qbp}, which makes the SM a widely accepted theory for particle physics. However, it has been widely recognized that the SM is just an effective theory of a much more fundamental one,  where the relevant energy scale far exceeds the detection capabilities of current experiments. The physics beyond the SM or the so called New Physics (NP), if discovered by the experiments, will provide critical clues for us to construct the fundamental theory.  In the passed decade, some signs of NP have been  observed by various of experimental groups~\cite{DayaBay:2018yms,T2K:2019bcf,Muong-2:2021ojo,Hays:2022qlw}. 

Heavy flavor physics offers one of the ideal platforms for searching NP.  Recently, some anomalies in heavy meson decays such as $R_{K^{(*)}}$~\cite{LHCb:2021trn} and $R_{D^{(*)}}$~\cite{Belle:2019rba,Paolucci:2022mpj} have been observed, which implies the existence of NP.  Besides the heavy mesons, nowadays the heavy baryon or especially the charm baryon decays have attracted the attention of the experiments~\cite{Xu:2022kkh}, and a number of charm baryon decay channels  have been measured by many experimental collaborations, such as Belle~\cite{Belle:2021mvw,Belle:2021vyq}, LHCb~\cite{LHCb:2022ouv} and BESIII~\cite{BESIII:2022bkj,BESIII:2022qaf}. 

In the latest two years, the puzzle about the branching fraction of the $\Xi_c^{0/+}\to\Xi^{-/0}\ell^+\nu$ decays emerges from deviation between the  experimental measurements and theoretical predictions:~\cite{Aliev:2021wat,Zhang:2021oja,Zhao:2021sje}. In our previous research~\cite{He:2021qnc}, the  branching fraction ${\cal B}(\Xi_c^0\to \Xi^-e^+\nu)$ from the SU(3) symmetry prediction has 6 $\sigma$ standard deviation from the experimental data~\cite{Belle:2021crz,ALICE:2021bli}.  Furthermore, not only the semi- but also the non-leptonic charm baryon decays have the SU(3) symmetry breaking effect~\cite{Zhong:2022exp}.  On the theoretical side, the $\Xi_c^{0/+}-\Xi^{\prime0/+}_c$ mixing effect is the most possible reason for explaining this puzzle~\cite{He:2021qnc,Ke:2022gxm,Geng:2022yxb,Liu:2022igi,Geng:2022xfz}.  However, on the experimental side, it is difficult  to search for such mixing directly since it is always emerged in the complex baryonic transitions.  Therefore,  finding  a suitable method for searching the $\Xi_c^{0/+}-\Xi^{\prime0/+}_c$ mixing effect is the main task of this work. 

In this work, we choose the four-body decay $\Xi_c\to\Xi^{(\prime)}(\Lambda \pi)\ell^+\nu$ as an ideal channel to search for the $\Xi_c^{0/+}-\Xi^{\prime0/+}_c$ mixing. Here $\Xi$ is a spin-1/2  octet state and $\Xi^\prime$ is a spin-3/2 decuplet state. In principle, this mixing  enables $\Xi_c^{0+}$ to decay into $\Xi^{\prime}$, so one should observe both the two resonants  $\Xi$ and $\Xi^{\prime}$ in this decay channel. However, the $\Xi_c\to\Xi^{\prime}(\Lambda \pi)\ell^+\nu$ is highly suppressed due to two reasons. The first one is that the $\Xi_c\to\Xi^{\prime}$ process is suppressed by ${\rm sin}{\theta_c}$ with $\theta_c$ being the mixing angle. The second reason is that the strong decay width of $\Xi^{\prime}\to \Lambda \pi$ is much smaller than that of $\Xi\to \Lambda \pi$. Therefore this channel can hardly be observed by the experiments. Instead of branching fraction, we propose the angular distributions of $\Xi_c\to\Xi^{(\prime)}(\Lambda \pi)\ell^+\nu$ to search for the $\Xi_c^{0/+}-\Xi^{\prime0/+}_c$ mixing. Note that $\Xi$ and $\Xi^{\prime}$ have different spins, which will lead to different angular distributions of the $\Lambda \pi$ states. It is possible for the experiments to distinguish the two resonants  $\Xi$ and $\Xi^{\prime}$ by distinguishing two exactly different angular distributions.

This paper is organized as follows. In Sec.II, we give the theoretical framework of this work, where  the helicity amplitudes for $\Xi_c\to\Xi^{(\prime)}(\Lambda \pi)\ell^+\nu$ decays are adopted to derive the angular distributions. In Sec.III, we give the angular distributions analysis of this process and put  forward an observable for searching the $\Xi_c^{0/+}-\Xi^{\prime0/+}_c$ mixing effect. In Sec.IV, the numerical results are performed using the form factors from Lattice calculation, Light-cone sum rules and light-front quark model so that to validate our analysis.  In the last section, a brief summary will be presented. Some calculation details are collected in the appendix.

\begin{figure*}[htp]
\includegraphics[width=1.3\columnwidth]{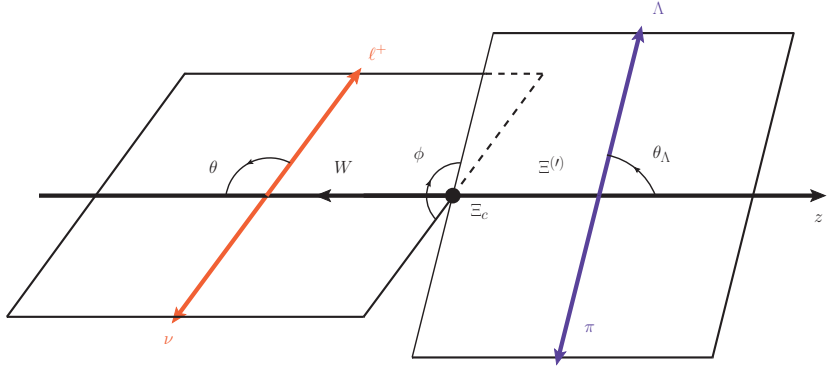} 
\caption{The kinematics for the  $\Xi^p_c\to\Xi^{(\prime)}(\Lambda \pi)\ell^+\nu$.  In the $\Xi^p_c$ baryon  rest frame, the  $\Xi^{(\prime)}$ moves along the $z$-axis. The $\theta(\theta_\Lambda)$ is defined as the angle between negative (positive) $z$-axis and the moving  direction of  $\ell^+$ ($\Lambda)$ in the $W$ ($\Xi^{(\prime)}$) rest frame. The $\phi$ is the angle between  the $\Xi^{(\prime)}$ and W cascade decay planes.}
\label{db}
\end{figure*}
\section{Helicity amplitude}
The deviation between theoretical prediction and experimental results can be explained by the $\Xi_c^{0/+}-\Xi^{\prime0/+}_c$ mixing~\cite{He:2021qnc,Geng:2022yxb}. After the mixing the physical state $(\Xi_c^{0/+p})$ is expressed as
 \begin{eqnarray}
\Xi_c^{0/+p}=\Xi_c^{0/+}\cos\theta_c+\Xi_c^{\prime0/+}\sin\theta_c,\label{mix}
\end{eqnarray}
where $\Xi^{(\prime)0+}_c$ are the flavor eigenstates. 
The physical state $\Xi^p_c$ can decay into decuplet baryon state $\Xi^\prime$ which is forbidden without the mixing effect. Therefore the processes $\Xi^p_c\to\Xi^\prime \ell^+\nu$ can be used to search for the mixing effect. However, as mentioned before it is difficult to measure this process directly because the $\Xi^p_c\to\Xi^\prime\ell^+\nu$ will suppressed by the factor  $\sin\theta_c$.

To search for the forbidden process $\Xi^p_c\to\Xi^\prime \ell^+\nu$, we will focus on the cascade decay process $\Xi^p_c\to\Xi^{(\prime)}(\Lambda \pi)\ell^+\nu$. Although the resonance $\Xi^\prime$ may have a tiny contribution which can be seemed as the systemic error. The angular distribution will provide additional information about the decay processes.

The  kinematics of $\Xi^p_c$ cascade decay are shown in Fig.~\ref{db}.  In the rest frame of the initial state $\Xi^p_c$,  the $\Xi^{(\prime)}$ moves along the z-axis. The angle $\phi$ is defined as the angle between the leptonic decay plane and $\Xi^{(\prime)}$ cascade decay plane, while $\theta(\theta_\Lambda)$ is the angle between the moveing direction of $\ell^+(\Lambda)$ and the positive (negative) direction of $z$-axis.


Using the Breit-Wigner form for the resonance, we can divide the amplitude of the cascade decay into several  Lorentz-invariant parts as
 \begin{eqnarray}
&&{\mathcal{M}}(\Xi^p_c\to\Xi^{(\prime)}(\Lambda \pi)\ell^+\nu)=\sum_{J_{\Xi^{(\prime)}}}\sum_{s_{\Xi^{(\prime)}}}{i\mathcal M}(\Xi^p_c\to\Xi^{(\prime)} \ell^+\nu)\notag\\
&&\qquad \quad \times \frac{i}{p_{\Xi}^2-m_{\Xi^{(\prime)}}^2+im_{\Xi^{(\prime)}}\Gamma_{\Xi^{(\prime)}}}{i\mathcal M}(\Xi^{(\prime)}\to\Lambda\pi),
\end{eqnarray}
with the momentum of resonance $p_\Xi^\mu=p_\Lambda^\mu+p_\pi^\mu$. 

For the $\Xi^p_c\to\Xi^{(\prime)} \ell^+\nu$ process, the relevant effective Hamiltonian is
\begin{eqnarray}
{\cal H}_{c\to s}&=&\frac{G_F}{\sqrt2} \left[V_{cs}^* \bar s \gamma^\mu(1-\gamma_5)c ~\bar \nu\gamma_\mu(1-\gamma_5) \ell\right] +h.c..
\end{eqnarray}
With the use of this effective Hamiltonian, the decay amplitude is written as the production of hadronic helicity amplitude and leptonic helicity amplitude:
\begin{eqnarray}
{i\mathcal M}(\Xi^p_c\to\Xi^{(\prime)} \ell^+\nu)&=&\sum_{s_w}\frac{G_F}{\sqrt2} V_{cs}^*\bar u_\nu \gamma_\rho(1-\gamma_5)\nu_\ell \epsilon^\rho(s_w)\notag\\
&&\times \langle \Xi^{(\prime)}| \bar s \gamma^\mu(1-\gamma_5)c |\Xi^p_c\rangle \epsilon^*_\mu(s_w)\notag\\
&=&\sum_{s_w}\frac{G_F}{\sqrt2} V_{cs}^*L^{s_w}_{s_\ell}(\phi,\theta)\times h^{s_{\Xi_c}}_{s_w,s_\Xi^{(\prime)}}
\end{eqnarray}

The Lorentz-invariant amplitude ${i\mathcal M}(\Xi^{(\prime)}\to\Lambda\pi)$ can be described by the  Wigner function and is parametrized as~\cite{Xing:2022uqu}
\begin{eqnarray}
{i\mathcal M}(\Xi^{(\prime)}\to\Lambda\pi)={\mathcal A^{(\prime)}}\times D^{J_\Xi}_{s_{\Xi^{(\prime)}},s_\Lambda}(\phi_\Xi,\theta_\Lambda),
\end{eqnarray}
where $J_\Xi$ is the total spin of $\Xi^{(\prime)}$ and $s_{\Xi^{(\prime)}}$, $s_\Lambda$ are the helicities of $\Xi^{(\prime)}$ and $\Lambda$ respectively. The $D^{J_\Xi}_{s_{\Xi^{(\prime)}},s_\Lambda}(\phi_\Xi,\theta_\Lambda)$ is the Wigner function~\cite{Workman:2022ynf} and $\phi_\Xi$ is the angle of $\Xi^{(\prime)}\Lambda\pi$ plane and x-z plane. The $\phi_\Xi$ is 0 in our work. The coefficient ${\mathcal A^{(\prime)}}$ can be determine by the decay width $\Gamma(\Xi^{(\prime)}\to\Lambda\pi)$:
\begin{eqnarray}
A&=&\sqrt{\Gamma(\Xi\to\Lambda\pi)8\pi m_\Xi^2/|p_\Lambda|}\notag\\
A^\prime&=&\sqrt{\Gamma(\Xi^\prime\to\Lambda\pi)16\pi m_{\Xi^{(\prime)}}^2/|p_\Lambda|}.\label{CA}
\end{eqnarray}
Then the total amplitude is expressed as
 \begin{eqnarray}
&&{\mathcal{M}}(\Xi^p_c\to\Xi^{(\prime)}(\Lambda \pi)\ell^+\nu)=\sum_{J_{\Xi^{(\prime)}}}\sum_{s_{\Xi^{(\prime)}},s_w} \frac{G_F}{\sqrt2} V_{cs}^* \notag\\
&&\qquad\qquad\times H^{J_{\Xi^{(\prime)}}}_{s_{\Xi_c},s_{\Xi^{(\prime)}}}L^{s_w}_{s_\ell}(\phi,\theta)D^{J_\Xi}_{s_{\Xi^{(\prime)}},s_\Lambda}(\phi_\Xi,\theta_\Lambda),\notag\\
&&H^{J_{\Xi^{(\prime)}}}_{s_{\Xi_c},s_{\Xi^{(\prime)}}}=\frac{i A^{(\prime)}}{p_{\Xi^{(\prime)}}^2-m_{\Xi^{(\prime)}}^2+im_{\Xi^{(\prime)}}\Gamma_{\Xi^{(\prime)}}}h^{s_{\Xi_c}}_{s_w,s_\Xi^{(\prime)}}\notag\\
&&\quad\qquad=L_{\Xi^{(\prime)}}h^{s_{\Xi_c}}_{s_w,s_\Xi^{(\prime)}}.
\end{eqnarray}
The differential decay width is expressed as~\cite{Huang:2021ots}: 
 \begin{eqnarray}
d\Gamma=d\Pi_4\times \frac{(2\pi)^4}{2m_{\Xi_c}}|{\mathcal M}(\Xi^p_c\to\Xi^{(\prime)}(\Lambda \pi)\ell^+\nu)|^2,
\end{eqnarray}
where $d\Pi_4$ is four body phase space integration.

\section{Angular distribution}

The differential decay width and other relevant observables of the four body decay $\Xi_c\to\Xi^{(\prime)}(\Lambda \pi)\ell^+\nu$ depend on the angle $\phi,\theta,\theta_\Lambda$ as shown in Fig.~\ref{db}. Now, expanding the Wigner function and the leptonic helicity amplitudes in the differential decay width, we arrive at 
\begin{widetext}
 \begin{eqnarray}
\frac{d\Gamma}{d\cos\theta d\cos\theta_\Lambda d\phi d p_\Xi^2 dq^2}&=&{\mathcal P}\bigg(L_{11}+L_{12}\cos\theta_\Lambda+L_{13}\cos2\theta_\Lambda +(L_{21}+L_{22}\cos\theta_\Lambda)\cos2\phi\notag\\
&&+(L_{31}+L_{32}\cos\theta_\Lambda+L_{33}\cos2\theta_\Lambda)\cos\theta\notag\\
&&+\big(L_{41}+L_{42}\cos2\phi+L_{43}\cos\theta_\Lambda+L_{44}\cos2\theta_\Lambda+L_{45}\cos2\theta_\Lambda\cos2\phi \big)\cos2\theta\notag\\
&&+(L_{51}\sin\theta_\Lambda+L_{52}\sin2\theta_\Lambda)\sin\theta\cos\phi+(L_{61}\sin\theta_\Lambda+L_{62}\sin2\theta_\Lambda)\sin2\theta\cos\phi\notag\\
&&+(L_{71}\sin\theta_\Lambda+L_{72}\sin2\theta_\Lambda)\sin\theta\sin\phi+(L_{81}\sin\theta_\Lambda+L_{82}\sin2\theta_\Lambda)\sin2\theta\sin\phi\notag\\
&&+(L_{91}+L_{92}\cos2\theta_\Lambda)\sin2\phi+(L_{101}+L_{102}\cos2\theta_\Lambda)\sin2\phi\cos2\theta\bigg),\notag\\
{\mathcal P}&=&\frac{G_F^2|V_{cs}|^2}{2}\frac{(1-\hat{m_\ell}^2)\sqrt{\lambda(m_{\Xi_c},\sqrt{p_\Xi^2},\sqrt{q^2})\lambda(\sqrt{p_\Xi^2},m_\Lambda,m_\pi)}}{(2\pi)^6 512 m^3_{\Xi_c} p^2_\Xi},\label{ad}
\end{eqnarray}
\end{widetext}
where  $q^\mu=p^\mu_{\Xi_c}-p_\Xi^\mu$ and $\hat{m_\ell}=\frac{m_\ell}{\sqrt{q^2}}$. $\lambda$ reads as 
 \begin{eqnarray}
\lambda(m_1,m_2,m_3)&=&\big((m_1+m_2)^2-m_3^2)\notag\\
&&\times((m_1-m_2)^2-m_3^2)\big).
\end{eqnarray}
The expressions of the coefficients $L_{ij}$ in Eq.~\eqref{ad} are given in Appendix.A. 

The resonance $\Xi^\prime$ can be detected by analyzing the shape of the differential decay width $d\Gamma/dp_\Xi^2$ as a function of $p^2_\Xi$, which reads as
 \begin{eqnarray}
\frac{d\Gamma}{d p_\Xi^2}&=&\frac{8\pi}{9}{\mathcal P}\bigg(9L_{11}-3L_{13}-3L_{41}+L_{44}\bigg).
\end{eqnarray}
This observable will have a peak around the $p^2_\Xi=m^2_{\Xi^\prime}$ according to the Breit-Wigner form of the resonance.  The contribution from each resonance is proportional to the branching fraction ${\cal B}(\Xi^{(\prime)}\to\Lambda\pi)$.
Therefore the process $\Xi^p_c\to\Xi^\prime\ell^+\nu$ is very difficult to be observed since the branching fraction ${\cal B}(\Xi^{\prime}\to\Lambda\pi)$ is extremely small. 

Besides the decay width the forward-backward asymmetry is another important observable. In this work, we define the normalized forward-backward asymmetry $A_{FB}$ as
\begin{eqnarray}
\frac{dA_{FB}}{dp_\Xi^2}&=&\frac{[\int^1_0-\int^0_{-1}]d\cos\theta_\Lambda\frac{d\Gamma}{dp^2_\Xi d\cos\theta_\Lambda}}{\int^1_{-1}d\cos\theta_\Lambda\frac{d\Gamma}{dp^2_\Xi d\cos\theta_\Lambda}}\notag\\
&=&\frac{3}{2}\frac{3L_{12}-L_{33}}{9L_{11}-3L_{13}-3L_{41}+L_{44}}\notag\\
&=&\frac{4}{3}\frac{\sum_{s_{\Xi_c},s_\Xi}{\mathcal R_e}(H^{\frac{1}{2}}_{s_{\Xi_c},s_{\Xi}}H^{\frac{3}{2}*}_{s_{\Xi_c},s_{\Xi}})}{\sum_{s_{\Xi_c},s_{\Xi^{(\prime)}}}(2|H^{\frac{1}{2}}_{s_{\Xi_c},s_{\Xi}}|^2+|H^{\frac{3}{2}}_{s_{\Xi_c},s_{\Xi^\prime}}|^2)}.\label{FAB}
\end{eqnarray}
where
 \begin{eqnarray}
&&\sum_{s_{\Xi_c},s_\Xi}{\mathcal R_e}(H^{\frac{1}{2}}_{s_{\Xi_c},s_{\Xi}}H^{\frac{3}{2}*}_{s_{\Xi_c},s_{\Xi}})\notag\\
&&\qquad=\frac{(p^2_\Xi-m^2_\Xi)(p^2_\Xi-m_{\Xi^\prime})-\Gamma_\Xi m_\Xi \Gamma_{\Xi^\prime} m_{\Xi^\prime}}{((p^2_\Xi-m^2_\Xi)^2+\Gamma_\Xi^2 m_\Xi^2)((p^2_{\Xi}-m^2_{\Xi^\prime})^2+\Gamma_{\Xi^\prime}^2 m_{\Xi^\prime}^2)}\notag\\
&&\qquad\times(\cos\theta_c h^{3,s_{\Xi_c}}_{s_w,s_\Xi}+\sin\theta_c h^{6,s_{\Xi_c}}_{s_w,s_\Xi})\sin\theta_c h^{6,s_{\Xi_c}}_{s_w,s_\Xi^{\prime}}.
\end{eqnarray}
$h^{3,s_{\Xi_c}}_{s_w,s_\Xi}$ and $h^{6,s_{\Xi_c}}_{s_w,s_\Xi^{(\prime)}}$ is the hadronic matrix element with the triplet charm baryon and sextet charm baryon as
  \begin{eqnarray}
&&h^{3, s_{\Xi_c}}_{s_w,s_\Xi}= \langle \Xi| \bar s \gamma^\mu(1-\gamma_5)c |\Xi_c\rangle \epsilon^*_\mu(s_w),\notag\\
&&h^{6, s_{\Xi_c}}_{s_w,s_\Xi^{(\prime)}}= \langle \Xi^{(\prime)}| \bar s \gamma^\mu(1-\gamma_5)c |\Xi^\prime_c\rangle \epsilon^*_\mu(s_w).
\end{eqnarray}
 It can be found that the forward-backward asymmetry is proportional to the interference of the amplitudes induced by the $\Xi$ and $\Xi^\prime$ resonants. Therefore, although the amplitudes induced by $\Xi^\prime$ are suppressed due to the tiny decay width of $\Xi^{\prime}\to\Lambda\pi$, $A_{FB}$ is still enhanced by the amplitudes induced by $\Xi$, which makes it possible to measure $A_{FB}$ by the experiments.  Furthermore, since the $A_{FB}$ obtained here is a function of $\theta_c$, we can determine this mixing angle as soon as the exact value of $A_{FB}$ is measured.  We will give the $\theta_c$ dependence of $A_{FB}$ in the Sec.~III.

On the other hand, as we discussed in Ref.~\cite{Xing:2022uqu}, it is possible to distinguish the resonants with different spins by $A_{FB}$. For the decay process of this work, the $dA_{FB}/dp^2_\Xi$ will have two zero points $s_1\sim m^2_\Xi,s_2\sim m^2_{\Xi^\prime}$ which can be obtained by solving the equation:
  \begin{eqnarray}
&&\sum_{s_{\Xi_c},s_\Xi}{\mathcal R_e}(H^{\frac{1}{2}}_{s_{\Xi_c},s_{\Xi}}H^{\frac{3}{2}*}_{s_{\Xi_c},s_{\Xi}})\notag\\
&&\qquad\propto\frac{(p^2_\Xi-m^2_\Xi)(p^2_\Xi-m_{\Xi^\prime})-\Gamma_\Xi m_\Xi \Gamma_{\Xi^\prime} m_{\Xi^\prime}}{((p^2_\Xi-m^2_\Xi)^2+\Gamma_\Xi^2 m_\Xi^2)((p^2_{\Xi}-m^2_{\Xi^\prime})^2+\Gamma_{\Xi^\prime}^2 m_{\Xi^\prime}^2)}\notag\\
&&\qquad=0.
\end{eqnarray}
The solutions read as
\begin{eqnarray}
s_1&=&\frac{1}{2}(m_\Xi^2+m_{\Xi^\prime}^2-\sqrt{(m_\Xi^2-m_{\Xi^\prime}^2)^2-4\Gamma_\Xi m_\Xi \Gamma_{\Xi^\prime} m_{\Xi^\prime}})\notag\\
&=&m_\Xi^2-\frac{\Gamma_\Xi m_\Xi \Gamma_{\Xi^\prime} m_{\Xi^\prime}}{m_{\Xi^\prime}^2-m_\Xi^2}+O(\Gamma_{\Xi^\prime})^2\notag,\\
s_2&=&\frac{1}{2}(m_\Xi^2+m_{\Xi^\prime}^2+\sqrt{(m_\Xi^2-m_{\Xi^\prime}^2)^2-4\Gamma_\Xi m_\Xi \Gamma_{\Xi^\prime} m_{\Xi^\prime}})\notag\\
&=&m_{\Xi^\prime}^2+\frac{\Gamma_\Xi m_\Xi \Gamma_{\Xi^\prime} m_{\Xi^\prime}}{m_{\Xi^\prime}^2-m_\Xi^2}+O(\Gamma_{\Xi^\prime})^2.
\end{eqnarray}
Note that $\Gamma_{\Xi^\prime(1530)}=0.0091{\rm GeV}$ is extremely small. Thus only using the leading term of each solution is precise enough for the following studies.  Now, $dA_{FB}/dp^2_\Xi$ has two zero points and each one is around the mass pole of  $\Xi$ or $\Xi^{\prime}$. This enables us to distinguish the two resonants and provides the evidence of $\Xi^{0/+}_c-\Xi_c^{0/+\prime}$ mixing.

\section{Numerical estimation}

\begin{figure*}[htp]
  \begin{minipage}[t]{0.48\linewidth}
  \centering
\includegraphics[width=0.85\columnwidth]{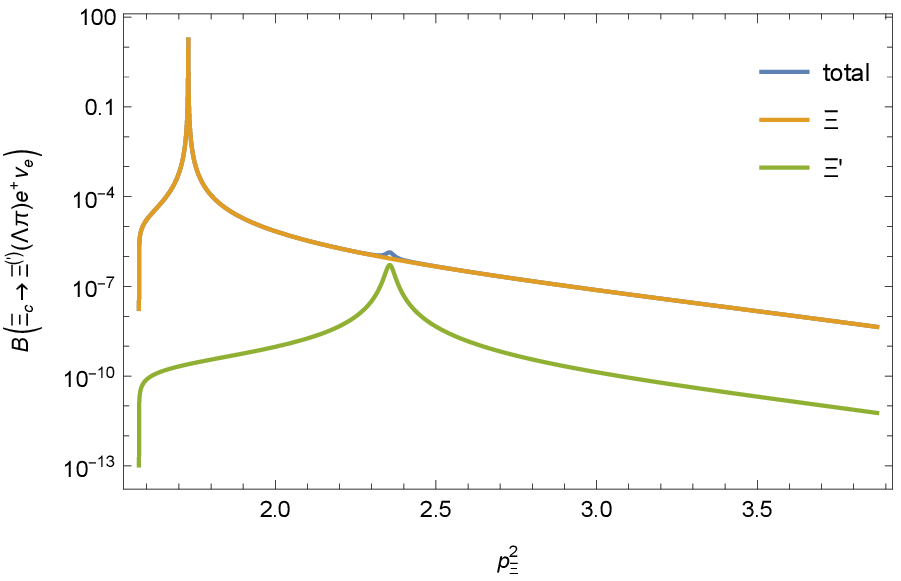} 
\end{minipage}
  \begin{minipage}[t]{0.48\linewidth}
  \centering
\includegraphics[width=0.85\columnwidth]{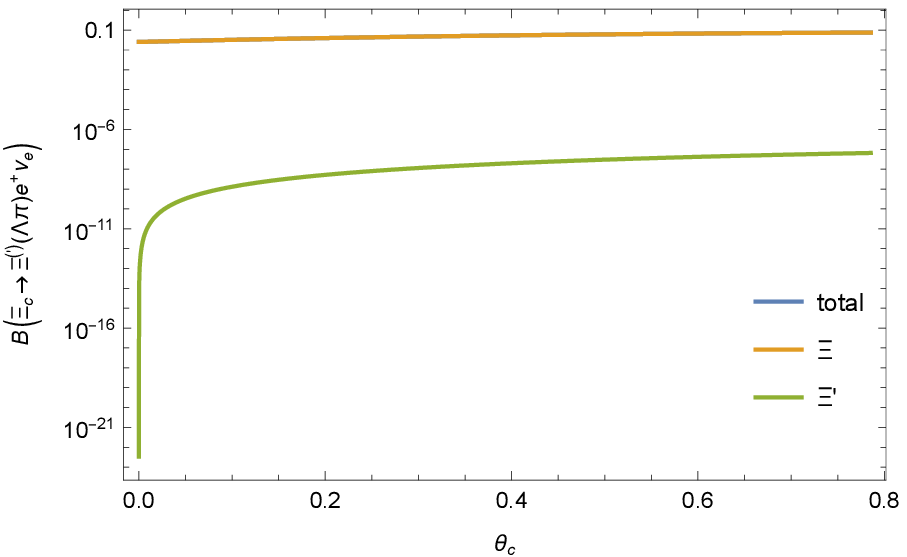} 
\end{minipage}
\caption{Branching fraction as functions of $p^2_\Xi$ (left) and the mixing angle (right). $\theta_c=0.137\pi$ is used for the left diagram.}
\label{p2d}
\end{figure*}

\begin{figure*}[htp]
  \begin{minipage}[t]{0.48\linewidth}
  \centering
\includegraphics[width=0.85\columnwidth]{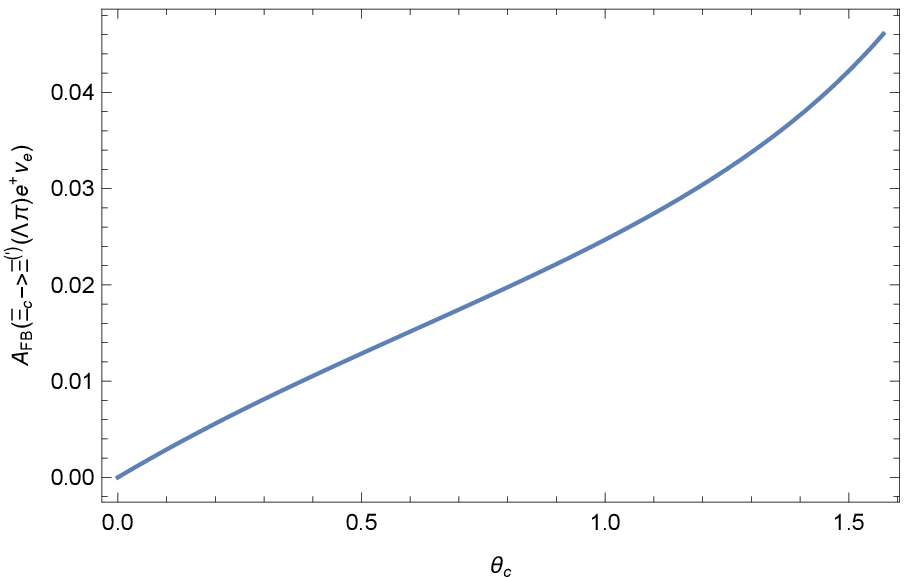} 
\end{minipage}
  \begin{minipage}[t]{0.48\linewidth}
  \centering
\includegraphics[width=0.85\columnwidth]{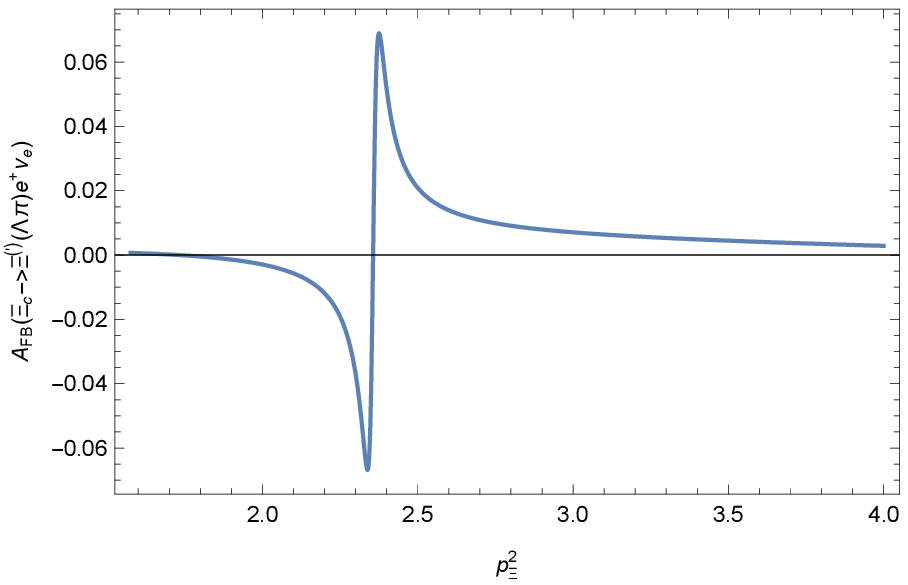} 
\end{minipage}
\caption{Forward-backward asymmetry as functions of the mixing angle (left) and $p^2_\Xi$ (right). $\theta_c=0.137\pi$ is used for the right diagram.}
\label{thafb}
\end{figure*}
In this section, we will give a numerical estimation by calculating the hadronic matrix element  $h^{s_{\Xi_c}}_{s_w,s_\Xi^{(\prime)}}$ with the form factors from Lattice results~\cite{Zhang:2021oja}, Light-cone Sum Rules~\cite{Azizi:2011mw} and quark model~\cite{Hsiao:2020gtc}. 
The hadronic matrix element is defined as
\begin{eqnarray}
&&h^{s_{\Xi_c}}_{s_w,s_\Xi^{(\prime)}}= \langle \Xi^{(\prime)}| \bar s \gamma^\mu(1-\gamma_5)c |\Xi^{p}_c\rangle \epsilon^*_\mu(s_w),
\end{eqnarray}
where the initial state $|\Xi^{p}_c\rangle$ is physical state which is the mixing state of the triplet and sextet charm baryon state in Eq.~\eqref{mix}. Then the matrix element can be expressed  as
\begin{eqnarray}
  \langle \Xi| \bar s \gamma^\mu(1-\gamma_5)c |\Xi^{p}_c\rangle &=&\cos\theta  \langle \Xi| \bar s \gamma^\mu(1-\gamma_5)c |\Xi_c\rangle \notag\\
&&+\sin\theta  \langle \Xi| \bar s \gamma^\mu(1-\gamma_5)c |\Xi^\prime_c\rangle \notag\\
 \langle \Xi^\prime| \bar s \gamma^\mu(1-\gamma_5)c |\Xi^{p}_c\rangle &=&\sin\theta  \langle \Xi^\prime| \bar s \gamma^\mu(1-\gamma_5)c |\Xi^\prime_c\rangle.
\end{eqnarray}
These hadronic  matrix element can be expressed by form factors as 
\begin{eqnarray}
&& \langle \Xi| \bar s \gamma^\mu(1-\gamma_5)c |\Xi_c\rangle=\notag\\
&&\quad \times \bigg(\bar u(p_{\Xi},s_{\Xi})[f_1\gamma^\mu+\frac{i\sigma^{\mu\nu}q_\nu}{m_{\Xi_c}}f_2+\frac{q^\mu}{m_{\Xi_c}}f_3]u(p_{\Xi_c},s_{\Xi_c})\notag\\
&&\quad -\bar u(p_{\Xi},s_{\Xi})[g_1\gamma^\mu+\frac{i\sigma^{\mu\nu}q_\nu}{m_{\Xi_c}}g_2+\frac{q^\mu}{m_{\Xi_c}}g_3]\gamma_5u(p_{\Xi_c},s_{\Xi_c})\bigg),\notag\\
&& \langle \Xi| \bar s \gamma^\mu(1-\gamma_5)c |\Xi^\prime_c\rangle=\notag\\
&&\quad \times \bigg(\bar u(p_{\Xi},s_{\Xi})[f^\prime_1\gamma^\mu+\frac{i\sigma^{\mu\nu}q_\nu}{m_{\Xi_c}}f^\prime_2+\frac{q^\mu}{m_{\Xi_c}}f^\prime_3]u(p_{\Xi_c},s_{\Xi_c})\notag\\
&&\quad -\bar u(p_{\Xi},s_{\Xi})[g^\prime_1\gamma^\mu+\frac{i\sigma^{\mu\nu}q_\nu}{m_{\Xi_c}}g^\prime_2+\frac{q^\mu}{m_{\Xi_c}}g^\prime_3]\gamma_5u(p_{\Xi_c},s_{\Xi_c})\bigg),\notag\\
&&\langle \Xi^\prime| \bar s \gamma^\mu(1-\gamma_5)c |\Xi^\prime_c\rangle =\bigg(\bar u_\rho(p_{\Xi},s_{\Xi^{\prime}})\bigg[\big(F_1\gamma^\mu+\frac{p^\mu_{\Xi_c}}{m_{\Xi_c}}F_2\notag\\
&&\quad+\frac{p_{\Xi^\prime}^\mu}{m_{\Xi^\prime}}F_3\big)\frac{p^\rho_{\Xi_c}}{m_{\Xi_c}}+g^{\mu\rho}F_4\bigg]\gamma_5 u(p_{\Xi_c},s_{\Xi_c})\notag\\
&&\quad-\bar u_\rho(p_{\Xi},s_{\Xi^{\prime}})\bigg[\big(G_1\gamma^\mu+\frac{p^\mu_{\Xi_c}}{m_{\Xi_c}}G_2+\frac{p_{\Xi^\prime}^\mu}{m_{\Xi^\prime}}G_3\big)\frac{p^\rho_{\Xi_c}}{m_{\Xi_c}}\notag\\
&&\quad+g^{\mu\rho}G_4\bigg]u(p_{\Xi_c},s_{\Xi_c})\bigg).
\end{eqnarray}
The form factors for $\Xi_c\to\Xi^\prime$ are taken from light-cone Sum Rule calculation~\cite{Azizi:2011mw}. Note that although this literature only calculated the form factors for $\Omega_c\to \Omega$, using SU(3) relations we can transform them to those for $\Xi_c\to\Xi^\prime$ by simply timing a factor  $\sqrt{2/3}$ on them~\cite{Geng:2017mxn}. The form factors for  $\Xi_c\to\Xi$ are taken from  Lattice QCD~\cite{Zhang:2021oja}.  In our calculation, the mixing angle is taken as $\theta_c=0.137\pi$ as given in Ref.~\cite{Geng:2022yxb}. The branching fraction ${\cal B}(\Xi^\prime\to \Lambda\pi)$ is temporarily set by a small value $0.001\%$ since it has not detected in experiment yet. In our analysis, we do not distinguish the processes $\Xi^0_c\to \Xi^{(\prime)-}(\Lambda\pi) e^+\nu$  and $\Xi^+_c\to \Xi^{(\prime)0}(\Lambda\pi) e^+\nu$. Therefore, it is suitable to give the numerical analysis for $\Xi^0_c\to \Xi^{(\prime)-}(\Lambda\pi )e^+\nu$ process as an example.

For the decay width,  the numerical results can be derived by integrating out the $p_\Xi^2,q^2$ and angle $\theta,\phi,\theta_\Lambda$ as
\begin{eqnarray}
&&\Gamma(\Xi^0_c\to \Xi^-(\Lambda\pi) e^+\nu)=2.476\times 10^{-13} {\rm GeV}\notag\\
&&\Gamma(\Xi^0_c\to \Xi^{\prime-}(\Lambda\pi) e^+\nu)=9.950\times 10^{-20} {\rm GeV}\notag\\
&&\Gamma(\Xi^0_c\to \Xi^{(\prime)-}(\Lambda\pi )e^+\nu)=2.476\times 10^{-13} {\rm GeV}\notag\\
&&{\cal B}(\Xi^0_c\to \Xi^{(\prime)-}(\Lambda\pi )e^+\nu)=5.68\%.
\end{eqnarray}
When the mixing angle $\theta_c$ become zero, the branching fraction of process will become ${\cal B}(\Xi^0_c\to \Xi^-(\Lambda\pi) e^+\nu)=2.4\%$. It is consistent with the previous work~\cite{Huang:2021ots}.
We also present the differential decay width $d\Gamma/dp^2_\Xi$ as a function of $p^2_\Xi$  in Fig.~\ref{p2d}. In this figure, one can find that the contribution of the $\Xi^\prime$ resonant is tiny.   Since the initial state $\Xi^p_c$ is the mixing state of the triplet and sextet charmed baryon, our result will also depend on the mixing angle. Therefore we also  present  the $\theta_c$ dependence in Fig.~\ref{p2d}.

For the forward-backward asymmetry, we can estimate its value by integrating out the $p^2_\Xi$ in Eq.~\eqref{FAB} and obtain
\begin{eqnarray}
&&A_{FB}(\Xi^0_c\to \Xi^-(\Lambda\pi) e^+\nu)=0.0158{\rm GeV^2}.
\end{eqnarray}
Since the $A_{FB}$ is sensitive to the mixing angle $\theta_c$, we can study the $\theta_c$ dependence of $A_{FB}$. The distribution of $A_{FB}(\theta_c)$ is shown in Fig.~\ref{thafb}.

One can easily find that the $A_{FB}$ is zero when the mixing angle vanishes $\theta_c=0$ and increases with the growing of $\theta_c$. Therefore, $\theta_c$ can be determined by measuring the $A_{FB}$ in the experiments. Furthermore, we have also shown the $p^2_\Xi$ dependence $A_{FB}$ in Fig.~\ref{thafb} by setting $\theta_c=0.137\pi$. The distribution of $dA_{FB}/dp^2_\Xi$ shows two zero points which are around the mass poles of $\Xi$ and $\Xi^\prime$ respectively, which is a strong signal of the $\Xi^{0/+}_c-\Xi_c^{\prime0/+}$ mixing effect and can be measured by  future experiments.

\section{Summary}
The angular distribution of $\Xi^p_c\to \Xi^{(\prime)}(\Lambda\pi)\ell^+\nu$ is analyzed in this work by introducing the $\Xi^{0/+}_c-\Xi_c^{\prime0/+}$ mixing effect.  Due to the $\Xi^{0/+}_c-\Xi_c^{\prime0/+}$ mixing effect, the physical state $\Xi^p_c$ can decay into $\Lambda\pi\ell^+\nu$ bypass  the resonance $\Xi^\prime$. Therefore the four body cascade decay process $\Xi_c\to \Xi^{(\prime)}(\Lambda\pi)\ell^+\nu$ becomes a good platform for searching the $\Xi^{0/+}_c-\Xi_c^{\prime0/+}$ mixing effect in the experiment.  

We have introduced an observable: forward-backward asymmetry $A_{FB}$, which can be used for searching mixing effect and measuring the mixing angle $\theta_c$.  Compared with the differential decay width, the advantage of $A_{FB}$ is that it can reflect both the $\Xi$ and $\Xi^\prime$ contributions, and has a good monotonous dependency on the mixing angle $\theta_c$. 

In our numerical estimation, we achieved  the value of decay width $\Gamma(\Xi^0_c\to \Xi^{(\prime)-}(\Lambda\pi )e^+\nu)=2.476\times 10^{-13} {\rm GeV}$ and branching fraction ${\cal B}(\Xi^0_c\to \Xi^{(\prime)-}(\Lambda\pi )e^+\nu)=5.68\%$. We find that the contribution of the $\Xi^\prime$ resonance is $4.019\times 10^{-5}\%$. We also estimate the value of integrated $A_{FB}$ as $A_{FB}(\Xi^0_c\to \Xi^{(\prime)-}(\Lambda\pi) e^+\nu)=0.0158{\rm GeV^2}$ and study  the $\theta_c$ and $p^2_\Xi$ dependence  of the $A_{FB}$. We believe our research will provide a useful guidance for searching $\Xi^{0/+}_c-\Xi_c^{\prime0/+}$ mixing effect in the future experiment.

\section{Acknowledgements}

We thank  Prof.  Xiao gang He and Prof. Wei Wang for useful discussions. 
This work  is supported in part by Natural Science Foundation of China under grant No. 12090064, 11735010, 11911530088, 12147147, by Natural Science Foundation of Shanghai under grant No. 15DZ2272100.

\appendix
  \section{Coefficient function in angular distribution}
  The specific expressions of coefficient $L_{ij}$ is
   \begin{widetext}
  \begin{eqnarray}
 L_{11}&=&-\frac{q^2}{8}(\hat{m}_\ell^2-1)\bigg(\hat{m}_\ell^2\big(3|H^{\frac{3}{2}}_{\frac{1}{2},\frac{3}{2}}|^2+16|H^{\frac{1}{2}}_{\frac{1}{2},\frac{1}{2}}|^2+10|H^{\frac{3}{2}}_{\frac{1}{2},\frac{1}{2}}|^2+5|H^{\frac{3}{2}}_{\frac{1}{2},-\frac{1}{2}}|^2+8|H^{\frac{1}{2}}_{\frac{1}{2},-\frac{1}{2}}|^2\big)+\notag\\
&&\big(9|H^{\frac{3}{2}}_{\frac{1}{2},\frac{3}{2}}|^2+16|H^{\frac{1}{2}}_{\frac{1}{2},\frac{1}{2}}|^2+10|H^{\frac{3}{2}}_{\frac{1}{2},\frac{1}{2}}|^2+15|H^{\frac{3}{2}}_{\frac{1}{2},-\frac{1}{2}}|^2+24|H^{\frac{1}{2}}_{\frac{1}{2},-\frac{1}{2}}|^2\big)\bigg)+\Big((s_{\Xi_c} ,s_\Xi)\to (-s_{\Xi_c} ,-s_\Xi)\Big),\notag\\
L_{12}&=&-2q^2(\hat{m}_\ell^2-1)\bigg(2(\hat{m}_\ell^2+1)\mathcal{R}_e(H^{\frac{3}{2}}_{\frac{1}{2},\frac{1}{2}}H^{\frac{1}{2}*}_{\frac{1}{2},\frac{1}{2}})+(\hat{m}_\ell^2+3)\mathcal{R}_e(H^{\frac{3}{2}}_{-\frac{1}{2},\frac{1}{2}}H^{\frac{1}{2}*}_{-\frac{1}{2},\frac{1}{2}})\bigg)+\Big((s_{\Xi_c} ,s_\Xi)\to (-s_{\Xi_c} ,-s_\Xi)\Big),\notag\\
L_{13}&=&\frac{3q^2}{8}(\hat{m}_\ell^2-1)\bigg(-(\hat{m}_\ell^2+1)2|H^{\frac{3}{2}}_{\frac{1}{2},\frac{1}{2}}|^2+(\hat{m}_\ell^2+3)(|H^{\frac{3}{2}}_{\frac{1}{2},-\frac{1}{2}}|^2-|H^{\frac{3}{2}}_{\frac{1}{2},\frac{3}{2}}|^2)\bigg)+\Big((s_{\Xi_c} ,s_\Xi)\to (-s_{\Xi_c} ,-s_\Xi)\Big),\notag\\
L_{21}&=&\frac{\sqrt{3}q^2}{4}(\hat{m}_\ell^2-1)^2\mathcal{R}_e(H^{\frac{3}{2}}_{\frac{1}{2},\frac{3}{2}}H^{\frac{3}{2}*}_{\frac{1}{2},-\frac{1}{2}})+\Big((s_{\Xi_c} ,s_\Xi)\to (-s_{\Xi_c} ,-s_\Xi)\Big),\quad L_{22}=-L_{21},\notag\\
L_{31}&=&-\frac{q^2}{2}(\hat{m}_\ell^2-1)\Big(3|H^{\frac{3}{2}}_{\frac{1}{2},\frac{3}{2}}|^2-5|H^{\frac{3}{2}}_{\frac{1}{2},-\frac{1}{2}}|^2-8|H^{\frac{1}{2}}_{\frac{1}{2},-\frac{1}{2}}|^2\Big)-\Big((s_{\Xi_c} ,s_\Xi)\to (-s_{\Xi_c} ,-s_\Xi)\Big),\notag\\
L_{32}&=&8q^2(\hat{m}_\ell^2-1)\bigg(\mathcal{R}_e(H^{\frac{3}{2}}_{-\frac{1}{2},\frac{1}{2}}H^{\frac{1}{2}*}_{-\frac{1}{2},\frac{1}{2}})\bigg)-\Big((s_{\Xi_c} ,s_\Xi)\to (-s_{\Xi_c} ,-s_\Xi)\Big),\notag\\
L_{33}&=&\frac{3q^2}{2}(\hat{m}_\ell^2-1)(|H^{\frac{3}{2}}_{\frac{1}{2},\frac{3}{2}}|^2+|H^{\frac{3}{2}}_{\frac{1}{2},-\frac{1}{2}}|^2)-\Big((s_{\Xi_c} ,s_\Xi)\to (-s_{\Xi_c} ,-s_\Xi)\Big),\notag\\
L_{41}&=&\frac{q^2}{8}(\hat{m}_\ell^2-1)^2\Big(3|H^{\frac{3}{2}}_{\frac{1}{2},\frac{3}{2}}|^2-10|H^{\frac{3}{2}}_{\frac{1}{2},\frac{1}{2}}|^2-16|H^{\frac{1}{2}}_{\frac{1}{2},\frac{1}{2}}|^2+5|H^{\frac{3}{2}}_{\frac{1}{2},-\frac{1}{2}}|^2+8|H^{\frac{1}{2}}_{\frac{1}{2},-\frac{1}{2}}|^2\Big)+\Big((s_{\Xi_c} ,s_\Xi)\to (-s_{\Xi_c} ,-s_\Xi)\Big),\notag\\
L_{42}&=&\frac{q^2}{4}(\hat{m}_\ell^2-1)^2\bigg(\sqrt{3}\mathcal{R}_e(H^{\frac{3}{2}}_{\frac{1}{2},\frac{3}{2}}H^{\frac{3}{2}*}_{\frac{1}{2},\frac{1}{2}})\bigg)+\Big((s_{\Xi_c} ,s_\Xi)\to (-s_{\Xi_c} ,-s_\Xi)\Big),\notag\\
L_{43}&=&-2q^2(\hat{m}_\ell^2-1)^2\bigg(2\mathcal{R}_e(H^{\frac{3}{2}}_{\frac{1}{2},\frac{1}{2}}H^{\frac{1}{2}*}_{\frac{1}{2},\frac{1}{2}})-\mathcal{R}_e(H^{\frac{3}{2}}_{\frac{1}{2},-\frac{1}{2}}H^{\frac{1}{2}*}_{\frac{1}{2},-\frac{1}{2}})\bigg)+\Big((s_{\Xi_c} ,s_\Xi)\to (-s_{\Xi_c} ,-s_\Xi)\Big),\notag\\
L_{44}&=&-\frac{3q^2}{8}(\hat{m}_\ell^2-1)^2(|H^{\frac{3}{2}}_{\frac{1}{2},\frac{3}{2}}|^2+2|H^{\frac{3}{2}}_{\frac{1}{2},\frac{1}{2}}|^2-|H^{\frac{3}{2}}_{\frac{1}{2},-\frac{1}{2}}|^2)+\Big((s_{\Xi_c} ,s_\Xi)\to (-s_{\Xi_c} ,-s_\Xi)\Big),\notag\\
L_{45}&=&-\frac{q^2}{4}(\hat{m}_\ell^2-1)^2\bigg(\sqrt{3}\mathcal{R}_e(H^{\frac{3}{2}}_{\frac{1}{2},\frac{3}{2}}H^{\frac{3}{2}*}_{\frac{1}{2},-\frac{1}{2}})\bigg)+\Big((s_{\Xi_c} ,s_\Xi)\to (-s_{\Xi_c} ,-s_\Xi)\Big),\notag\\
L_{51}&=&-q^22\sqrt{2}(\hat{m}_\ell^2-1)\bigg(\sqrt{3}\mathcal{R}_e(H^{\frac{3}{2}}_{\frac{1}{2},\frac{3}{2}}H^{\frac{1}{2}*}_{\frac{1}{2},\frac{1}{2}})-\mathcal{R}_e(H^{\frac{3}{2}}_{\frac{1}{2},-\frac{1}{2}}H^{\frac{1}{2}*}_{\frac{1}{2},\frac{1}{2}})+\mathcal{R}_e(H^{\frac{3}{2}}_{\frac{1}{2},\frac{1}{2}}H^{\frac{1}{2}*}_{\frac{1}{2},-\frac{1}{2}})\bigg)-\notag\\
&&\Big((s_{\Xi_c} ,s_\Xi)\to (-s_{\Xi_c} ,-s_\Xi)\Big),\notag\\
L_{52}&=&-q^2\sqrt{6}(\hat{m}_\ell^2-1)^2\mathcal{R}_e(H^{\frac{3}{2}}_{\frac{1}{2},\frac{3}{2}}H^{\frac{1}{2}*}_{\frac{1}{2},\frac{1}{2}})-\Big((s_{\Xi_c} ,s_\Xi)\to (-s_{\Xi_c} ,-s_\Xi)\Big),\notag\\
L_{61}&=&q^2\sqrt{2}(\hat{m}_\ell^2-1)^2\bigg(-\sqrt{3}\mathcal{R}_e(H^{\frac{3}{2}}_{\frac{1}{2},\frac{3}{2}}H^{\frac{1}{2}*}_{\frac{1}{2},\frac{1}{2}})-\mathcal{R}_e(H^{\frac{3}{2}}_{\frac{1}{2},-\frac{1}{2}}H^{\frac{1}{2}*}_{\frac{1}{2},\frac{1}{2}})+\mathcal{R}_e(H^{\frac{3}{2}}_{\frac{1}{2},\frac{1}{2}}H^{\frac{1}{2}*}_{\frac{1}{2},-\frac{1}{2}})\bigg)-\notag\\
&&\Big((s_{\Xi_c} ,s_\Xi)\to (-s_{\Xi_c} ,-s_\Xi)\Big),\notag\\
L_{62}&=&q^2\sqrt{\frac{3}{2}}(\hat{m}_\ell^2-1)^2\mathcal{R}_e(H^{\frac{3}{2}}_{\frac{1}{2},\frac{3}{2}}H^{\frac{1}{2}*}_{\frac{1}{2},\frac{1}{2}})+\Big((s_{\Xi_c} ,s_\Xi)\to (-s_{\Xi_c} ,-s_\Xi)\Big),\notag\\
L_{71}&=&q^22\sqrt{2}(\hat{m}_\ell^2-1)^2\Big(\mathcal{I}_m(H^{\frac{3}{2}}_{\frac{1}{2},-\frac{1}{2}}H^{\frac{1}{2}*}_{\frac{1}{2},\frac{1}{2}})+\mathcal{I}_m(H^{\frac{3}{2}}_{\frac{1}{2},\frac{1}{2}}H^{\frac{1}{2}*}_{\frac{1}{2},-\frac{1}{2}})+\sqrt{3}\mathcal{I}_m(H^{\frac{3}{2}}_{\frac{1}{2},\frac{3}{2}}H^{\frac{1}{2}*}_{\frac{1}{2},\frac{1}{2}})\Big)+\Big((s_{\Xi_c} ,s_\Xi)\to (-s_{\Xi_c} ,-s_\Xi)\Big),\notag\\
L_{72}&=&q^2\sqrt{6}(\hat{m}_\ell^2-1)^2\mathcal{I}_m(H^{\frac{3}{2}}_{\frac{1}{2},\frac{1}{2}}H^{\frac{3}{2}*}_{\frac{1}{2},\frac{3}{2}})+\Big((s_{\Xi_c} ,s_\Xi)\to (-s_{\Xi_c} ,-s_\Xi)\Big),\notag\\
L_{81}&=&-q^2\sqrt{2}(\hat{m}_\ell^2-1)^2\Big(\mathcal{I}_m(H^{\frac{3}{2}}_{\frac{1}{2},-\frac{1}{2}}H^{\frac{1}{2}*}_{\frac{1}{2},\frac{1}{2}})+\mathcal{I}_m(H^{\frac{3}{2}}_{\frac{1}{2},\frac{1}{2}}H^{\frac{1}{2}*}_{\frac{1}{2},-\frac{1}{2}})-\sqrt{3}\mathcal{I}_m(H^{\frac{3}{2}}_{\frac{1}{2},\frac{3}{2}}H^{\frac{1}{2}*}_{\frac{1}{2},\frac{1}{2}})\Big)-\Big((s_{\Xi_c} ,s_\Xi)\to (-s_{\Xi_c} ,-s_\Xi)\Big),\notag
  \end{eqnarray}
    \begin{eqnarray}
L_{82}&=&-q2\sqrt{\frac{3}{2}}(\hat{m}_\ell^2-1)^2\mathcal{I}_m(H^{\frac{3}{2}}_{\frac{1}{2},\frac{1}{2}}H^{\frac{3}{2}*}_{\frac{1}{2},\frac{3}{2}})+\Big((s_{\Xi_c} ,s_\Xi)\to (-s_{\Xi_c} ,-s_\Xi)\Big),\quad L_{91}=-\frac{1}{2\sqrt{2}}L_{82},\quad L_{92}=-L_{91},\notag\\
L_{101}&=&L_{92},\quad L_{102}=L_{91}.
  \end{eqnarray}
   \end{widetext}

\end{document}